\documentclass[aps,twocolumn,prl,tighten,flushrt]{revtex4}
\usepackage{graphicx}

\input epsf

\def\be{\begin{equation}}
\def\ee{\end{equation}}
\def\bea{\begin{eqnarray}}
\def\eea{\end{eqnarray}}

\def\cmm2{{\,\rm cm^{-2}}}
\def\cm2{{\,{\rm cm}^2}}
\def\cmm3{{\,{\rm cm}^{-3}}}
\def\gcmm3{{\,{\rm g\,cm^{-3}}}}

\def\fun#1#2{\lower3.6pt\vbox{\baselineskip0pt\lineskip.9pt
  \ialign{$\mathsurround=0pt#1\hfil##\hfil$\crcr#2\crcr\sim\crcr}}}

\def\fm{{\rm~fm}}

\def\eg{{e.g., }}

\def\p3m{P$^3$M}

\def\fun#1#2{\lower3.6pt\vbox{\baselineskip0pt\lineskip.9pt
  \ialign{$\mathsurround=0pt#1\hfil##\hfil$\crcr#2\crcr\sim\crcr}}}
\newcommand{\het}{$^3$He~}
\newcommand{\hef}{$^4$He~}
\newcommand{\lisi}{$^6$Li~}
\newcommand{\lise}{$^7$Li~}
\newcommand{\bese}{$^7$Be~}

\begin{document}
\bibliographystyle{prsty}
\title{Big Bang Nucleosynthesis with Bound States of Long-lived
  Charged Particles}
\author{Manoj\ Kaplinghat and Arvind\ Rajaraman}
\affiliation{Department of Physics and Astronomy\\
University of California, Irvine, California 92697, USA}
\date{\today}

\begin{abstract}
Charged particles (X) decaying after primordial nucleosynthesis are
constrained by the requirement that their decay products should not
change the light element abundances drastically. If the decaying
particle is negatively charged (X$^-$) then it will bind to the
nuclei. We consider the effects of the decay of X when bound to
Helium-4 and show that this will modify the Lithium abundances.
\end{abstract}
 \pacs{98.70.Vc} \maketitle

\section{Introduction}

Many models of physics beyond the standard model have long-lived
charged particles  in their spectrum. For example, in the class of
supergravity models  where the gravitino is 
the lightest supersymmetric particle (LSP), the next lightest (NLSP)
can be a long-lived charged slepton \cite{feng03a}. Similar
phenomenology arises in Universal Extra Dimension models
\cite{feng03c}. Another example  
is the class of SUSY models with axino dark matter where again the
NLSP could be long-lived \cite{covi01,covi04,brandenburg05}. A third
example is a supersymmetric model with almost degenerate LSP and NLSP
such that the dominant decay channel for the NLSP is kinematically
suppressed \cite{sigurdson03,profumo04}.

Here we consider a new aspect of the early universe cosmology of
these particles that affects the light element abundances. Other
effects of these charged particles were previously considered in
 \cite{dimopoulos89,kawasaki94,khlopov94,sedelnikov95,holtmann96,holtmann98,jedamzik99,cyburt03,feng03b,jedamzik04,kawasaki04,jedamzik06a,jedamzik06b,steffen06} 

If the particle is negatively charged, it  will eventually form a
bound state with the positively charged nuclei.
When the particle decays, the decay products will electromagnetically
interact with the nearby nucleus. If the nucleus is $^4$He, it may
break up,
creating $^3$He, T, D, p, n. These $^3$He, T and D nuclei will find
other \hef nuclei and  interact to form \lisi and \lise.

In addition to the production mechanism described above, charged
particle decays result in a large non-thermal photon background that
can destroy the created \lisi and \lise nuclei.
When both the production and destruction processes are taken into
account, we find that the
overall \lisi production could be in the same range as the
Lithium isotopic abundance measured in few low metallicity popII stars 
\cite{smith93,smith98,nissen99,vangioni-flam99,nissen00,asplund05}.

Other effects can come from the elastic scattering of the charged
particle produced in the decay from the \hef nucleus, and also from
the change in the motion of the \hef  in the bound state. We find that
these effects produce insignificant changes to standard BBN
abundances. 

In this work we only consider singly charged bound state of \hef
nuclei \cite{cahn81,belotsky05,fargion05,khlopov06}. We ignore the
production of neutral atoms of \hef bound to two X. We also do not
consider the bound state of protons. This binding process starts below
about 0.5 keV and could be important for lifetimes larger than about a
year.  Another interesting process we have not considered here is that
of forming stable ${\rm Be}^8$-X bound states \cite{cahn81}. A stable
${\rm Be}^8$-X bound state has important implications for the
production of Carbon in the early universe. We hope to return to these 
issues in future work. 

While this work was being completed, two related articles appeared on
the arXiv. The three articles all consider different processes and
hence complement each other. The first by
Pospelov \cite{pospelov06} pointed out that 
the nuclear processes with a photon in the final state will be
modified by the presence of the strong electric field of the X
particle. Pospelov showed that the enhanced D on X-\hef cross-section
could produce orders of magnitude larger \lisi abundance. However, it
is important to note that if the decay lifetime is about $10^6$
seconds or larger, then \lisi will also be destroyed due to the 
non-thermal photons from the decay.

The second paper by Kohri and Takayama \cite{kohri06} worked out
the changes in the light element abundances due to the change in
coulomb barrier in the presence of the $X^-$, and due to the changes
in the kinematics and energy balance for the reactions. They found
that the Lithium abundances can be significantly modified from
the standard BBN predictions. 

\section{Rates}

To calculate the magnitude of the effects we mentioned in the
previous section, we need to calculate the following quantities.
\begin{enumerate}
\item $f_b$, the fraction of Helium-4 nuclei bound to the charged
  particles.
\item $f_P$, the fraction of the bound \hef nuclei which are
destroyed
  by photodisintegration.
\item $f_3$, the average fractions of \het and T produced by the
photodisintegration of a \hef nucleus. \item $f_L$, the fraction of
\het and T that formed in the
  photo-disintegration that find another \hef nucleus and form \lisi.
\end{enumerate}

In what follows we will concentrate on the \lisi abundance. Some
\lise is also produced through ${}^4{\rm He}({}^3{\rm He},\gamma){}^7{\rm
Be}$ and ${}^4{\rm He}({\rm T},\gamma){}^7{\rm Li}$. However, these
reactions have an electromagnetic component and hence are slower than
the \lisi production reactions.

The total fraction of \lisi to Hydrogen (by number) is then
$f_bf_Pf_3f_LY_P/4/(1-Y_P/2)$ where $Y_P$ is the total \hef mass
fraction (bound or not) in the absence of the decay. The effects on
the abundances of the other nuclei can also be similarly determined.

\section{Charged particle recombination}

\begin{figure}[t]
\includegraphics[width=\columnwidth]{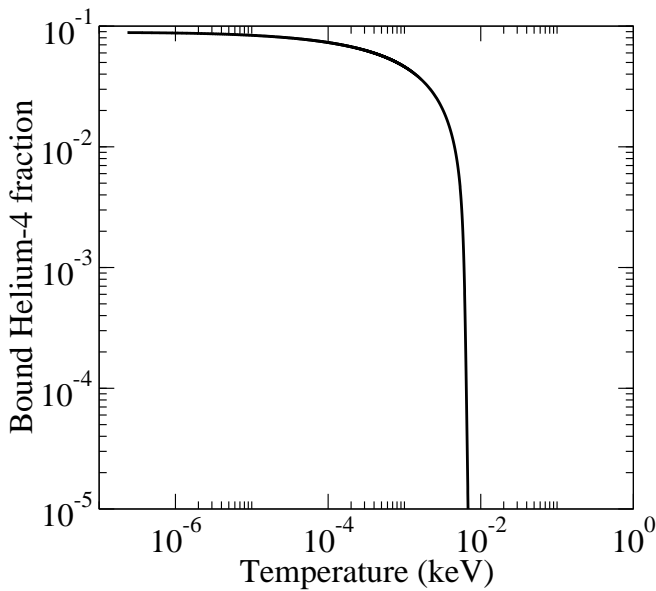}
    \caption{\label{fig:bound}
      Solid curve is the fraction of helium nuclei that are bound to
      a long-lived massive charged particle. The number density of
      these charged particles is set by the requirement that the
      massive neutral particle it decays into be all the dark matter
      in the universe and the curve here is for a dark matter particle
      mass of 200 GeV. We have taken the decay lifetime to be much
      larger than the age of the universe at the temperatures shown
      above.  
}
\end{figure}

 The bulk of \hef forms at $T\equiv T_{\alpha,f}
\simeq 70 keV$. If the decay happens significantly before this
temperature $T_{\alpha,f}$, then the effects we are interested in
here are not important. For the rest of the discussion we will assume
that the decay lifetime is larger than 
$1/H(T_{\alpha,f})$. 

We also assume that the predominant component of X is negatively
charged. Thus we are guaranteed that X will bind to the positively
charged nuclei.

The equation for the evolution of the fraction of charged X particles
that are not bound ($x_f$) is given by
\begin{equation}\label{eq:recomb}
\dot{x}_f=\left[ 
\beta (1-x_f)
- Rx_fn_B{1\over 4}Y_p(t)(1-f_b)\right] \,.
\end{equation}
In the above equation, $R$ is the recombination rate and $\beta$ is
the ionization rate. The recombination rate may be calculated in a
manner analogous to the recombination of Hydrogen to yield
$R=5(\alpha/m)\sqrt({\rm BE}/T)\ln({\rm BE}/T)$. The ionization
cross-section and hence $\beta$ can then be found from the usual
analysis of the equilibrium case.

\section{Photo-disintegration of Helium-4}

The photo-disintegration occurs through a one-photon exchange
diagram, where the charged decay product emits a photon of invariant
momentum $q^2=-Q^2$, which hits the Helium nucleus and breaks it up.

Photo-disintegration can occur as long as the energy transferred to
the hadronic system, is larger than the binding energy, which is here
28.8 MeV. Unfortunately, the cross-section for the process
$\gamma+He^4\rightarrow {any}$ has not been measured for all the
energies that we need. For large $Q^2$, one can use the formulae for
deep inelastic scattering, but these are only valid when $Q^2$ is
greater than $Q_0^2\sim 1GeV^2$.

  To proceed, we will restrict our analysis to photon momenta
  satisfying $Q^2>Q_0^2$. This will result in a conservative
  estimate of the cross-section; the true cross-section will
  certainly
receive additional contributions from photons below this cutoff.

For $Q^2>Q_0^2$, the cross section is given by
\begin{eqnarray}
{d\sigma_D\over dQ^2} & = & {2\pi \alpha^2\over Q^4}\int_0^1dx \nonumber\\
& & \times \sum_ff_f(x)Q_f^2[1+(1-{Q^2\over xs})^2]\theta(xs-Q^2) \,.
\end{eqnarray}

Here $f_f(x)$ is the parton distribution function of flavor $f$ at
longitudinal fraction $x$. $Q_f$ is the corresponding charge. The
final factor above is a kinematic constraint.

Unlike the usual situation in deep inelastic scattering, here we
should integrate over the allowed $Q^2$. For a fixed $x$ the range is
$Q_0^2<Q^2<xs$. The allowed range of $x$ is then between ${Q_0^2\over
s}$ and 1.

We then have to perform an integration over $x$. This is hampered by
our lack of knowledge of $f_f(x)$. Qualitatively, $f_f(x)$ increases
at smaller $x$, so this integral is expected to be dominated by the
smaller values of $x$.

We will assume that $\epsilon={Q_0^2\over s}\ll 0.1$, and restrict
the range of the integral between $x=\epsilon$ and $x=0.1$. Over this
range, we can  take approximately $F_2(x)=xf_f(x)Q_f^2\sim 1$
\cite{arneodo96}.

With these approximations, we find \bea \sigma\sim
-{2\pi\alpha^2\over Q_0^2}ln(10Q_0^2/s)\,. \eea

Using $1GeV^{-1}\sim 0.2 fm$,  the cross section evaluates to about
$2\pi\times 10^{-4}\times 0.04\times 10^{-26}cm^2 ln(10Q_0^2/s)\sim
0.3\times 10^{-3} ln(10Q_0^2/s)  mb$ in almost exact agreement with a
calculation from GEANT.

To derive a probability from this quantity, we must multiply by the
flux, and for this we must examine the geometry. The helium nucleus
orbits around X at a distance given by $d \simeq m^{-1}/Z\alpha =
3.6\fm$ where we have taken $m$ to be the mass of the helium nucleus.
The size of the helium nucleus may be taken to be
$r=1.44A^{1/3}\fm=2.3\fm$. The solid angle subtended by the nucleus
at the position of X is $2\pi(1-\cos\phi)$ where
$\cos^2\phi=1-r^2/d^2$ and the probability of disassociation is
\begin{equation}
{\sigma_D \over 4\pi d^2} {2(1-\cos\phi)\over \sin^2\phi\cos^2\phi}
\simeq 0.012\sigma_D\fm^{-2}\,.
\end{equation}
The probability of an interaction between the emitted particle and
the nucleus is thus $0.012 \times 3 \times 10^{-5} = 3.6 \times
10^{-7}$ for a cross-section of $3 \times 10^{-4}$mb.

\section{Nuclei from the photo-disintegration}

The photo-disintegration of \hef described in the previous section
typically breaks up the nucleus. The nuclei will split into lighter
elements $^3$He, T, D, p and n. The exclusive cross-sections for
these processes are unknown. We will assume 
following \cite{protheroe94}
that the helium nucleus is always destroyed, and produces \het and T
with the cross-sections $\sigma_3=\sigma_D={4\over 9}\sigma_P$. In
other words, the photodisintegration produces a \het or T nucleus in
about half the interactions, and D in about half the  interactions.
We will assume that \het and T are produced in equal amounts.

\section{Production of Lithium-6}

The \het in the photo-dissociation of \hef are extremely
relativistic. They can then scatter off background \hef nuclei, and
produce Lithium.  The 
rate for this is given by $\Gamma_\alpha=3.6\times 10^{-15} Y_P
T_{\rm MeV}^3 (\sigma/40{\rm mb}) (\eta_{10}/6) {\rm
  MeV}$ where $\sigma$ is the appropriate nuclear cross-section.
  In comparison, the expansion rate of
the universe is  $H(T)=2.5\times 10^{-22} T_{\rm MeV}^2 {\rm MeV}$.
Comparing the two rates, we see that the nuclear reaction rate will
be larger than the expansion rate of the universe until $T \sim 0.3
{\rm eV} (40 {\rm mb}/<\sigma v>)$.

The nuclear reaction rate that we have quoted above assumes that the
\het are energetic enough to overcome the Coulomb barrier. This is
certainly true for the nuclei at the moment of decay. 
However, these nuclei will lose energy due to a variety of processes
\cite{kawasaki94,kawasaki95},  the most important of which  are 
the energy losses due to Coulomb and Compton scattering.

The time scale for these energy loss processes is much shorter than
the nuclear reaction rate as well as the expansion rate. As a
consequence one may write the probability of \lisi to form as
\cite{dimopoulos89}
\begin{equation}
f_L=\int_{E_{\rm th}}^{K_0} dK \sigma_{3+4}(K) v {Y_P \over 4}
n_B(T) \left({dK \over dT}\right)^{-1}
\end{equation}
where $v$ is the relative speed, $K$ is the kinetic energy of the
nucleon, $dK/dT$ is the total energy loss rate and $\sigma_{3+4}$
is the cross-section for \het on \hef reaction. $E_{\rm th}$ is the
minimum energy required for the reaction to occur. The upper limit to
the kinetic energy $K_0$ is the energy with which the \het nucleus is
born. As it turns out, the integral is insensitive to the upper limit
as long as it is larger than about 100 MeV. This is due to the fast
energy degradation rate.

This is not the end of the story. When the decay occurs the charged
daughter particle carries off a large amount of energy. All of this
gets converted to non-thermal photons \cite{kawasaki95}. In the time
it takes for these photons to thermalize, they can find a \lisi (and
other nuclei) and photo-dissociate it if the energy is sufficient to
cross the threshold.

We can calculate the non-thermal distribution of the photons
following \cite{dimopoulos89,kawasaki95,cyburt03}. The initial energy
injection is rapidly brought to a quasi-static equilibrium state by
pair production off of the background (thermal) photons and inverse
Compton scattering. On a longer timescale (but still shorter than the
age) processes like photon-photon scattering and Compton scattering
redistribute the energy and lead to thermalization. 
We assume that the thermalization is mainly due to Compton scattering
and calculate the differential non-thermal photon number density
following \cite{cyburt03}.

With the expression for the non-thermal photon distribution at hand,
one may write the rate for \lisi production as
\begin{eqnarray}
{d\over d\ln(T)}Y_6(T) &=& -{n_B(T)  \over \tau H(T)}
f_L(T)f_3f_Pf_b(T)\nonumber\\
&+&Y_6(T){\Gamma_{6,\gamma}\over
  H(T)}\,
\end{eqnarray}
where subscript ``6'' stands for \lisi and $\Gamma_{6,\gamma}$ is the
total photo-dissociation rate for $\gamma+$\lisi
$\rightarrow$ anything. $Y$ denotes
the number density relative to the number density of baryons.

Similarly, the interaction of \het and T with the background protons
and photons can degrade the nuclear energy.
We have checked that these are negligible effects for the temperatures 
of around 10 keV and lower that we are interested in.

The non-thermal photons may also dissociate the \het before they can
collide with \hef nuclei. To estimate this, we note that the minimum
threshold for \het destruction is about 5.5 MeV. In order to produce
photons with this energy, one would have to be at a temperature lower
than about 2 keV.
Thus, this could only be a problem for long lifetimes. Also, the
spectrum at these energies is falling sharply
\cite{kawasaki95,cyburt03}. We estimate that this effect cannot be
significant unless decay lifetimes are larger than $10^6$ seconds and
energy released is larger than TeV. 

\begin{figure}[t]
\includegraphics[width=\columnwidth]{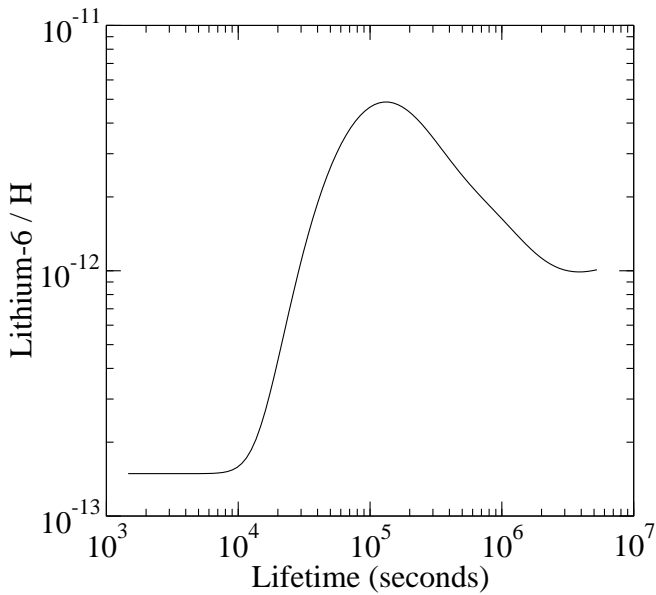}
    \caption{\label{fig:li6}
      Curve shows the ratio of number density of \lisi to that of
      H produced due to the process described in the text. The
      model assumed to make this plot is a gravitino Super-WIMP with
      mass of 100 GeV and a heavier charged slepton like stau. We
      start with an initial ${}^6{\rm Li}/{\rm H}$ ratio of $1.4
      \times 10^{-13}$.
}
\end{figure}

The bottom-line from the above calculation is that it is possible to
produce \lisi of order $10^{-12}$ (relative to the number density of
H) in large regions of decaying charged particle model parameter
space. This is clear from the Figure \ref{fig:li6} where we have
assumed that a slepton is decaying to a gravitino and lepton, with the  
gravitino having a mass of 100 GeV. As the mass of the gravitino is
lowered, the amount of \lisi produced decreases. Note that the masses
where a reasonable amount of \lisi is produced corresponds to
astrophysically interesting lifetimes of about a month
\cite{kaplinghat05,cembranos05}. We note in closing that the
dissociation by non-thermal photons will be insignificant for models
where the mass of the decaying and daughter particles is fine-tuned to
be small \cite{sigurdson03}. 

\section{Production of Lithium-7 and other reactions}

After the decay, all the \hef nuclei have the orbital kinetic energy
$\sim 0.16$ MeV.
The kinetic energy is still not enough to initiate any endothermic
nuclear reactions. We therefore consider the following processes
${}^3{\rm He}(\alpha,\gamma){}^7{\rm Be}$ and ${}^3{\rm
  H}(\alpha,\gamma){}^7{\rm Li}$. Proceeding as in the \lisi
production case, we find that this process produces insignificant
amounts of \lise and \bese. 

The  \hef nuclei could also receive a large kick due to elastic
scattering off of the decay produced charged particle.  
Our calculations show that for small energy releases of order 10 GeV,
it is possible to produce \lise in quantities comparable to the
standard BBN yield. However, in concrete cases like the Super-WIMP
scenario \cite{feng03a,feng03b}, the lifetime for such small energy
releases is large which means  that the destruction rate due to
non-thermal photons is large. The destruction effect dominates and
significant amounts of \lise are not produced.

We also note that the break-up of \hef produces energetic neutrons
and protons that in turn can break up the background \hef nuclei. 
However, we have seen that only about 1 in a million bound \hef is
broken up and hence the destruction due to these energetic protons
and neutrons is negligible.

\section{Discussion}
We have concentrated here on the collision of \het with background
\hef nuclei that would produce \lisi nuclei. Standard BBN does not
produce much \lisi and thus the expectation is that this could be an
important probe. The SBBN yield (without decays) is about $10^{-13}$
compared to the number density of H. This should be compared to the
measured lithium isotopic ratios in some low metallicity popII stars
which (conservatively) range between 0.01 and 0.1. A reasonable
constraint on \lisi then is that its abundance should not be much
larger than that of \lise nuclei (about $10^{-10}$ relative to the H
number density).  It is of course possible that the \lisi primordial
abundance is larger, but then we are left with the question of why
\lisi  is depleted in much larger amounts than \lise by astrophysical 
mechanisms. 

We found that in some regions of parameter space it is possible to
produce \lisi in quantities that match the observed ``plateau''
abundance. This does not imply that the decaying charged particle
scenario can explain the \lisi abundance. A full calculation including
all the light elements would be necessary to ascertain that. We also
note that there are astrophysical scenarios wherein the observed
abundance of \lisi finds an explanation (\eg
\cite{suzuki02,lambert04,rollinde05,reeves05,prantzos06,prodanovic06,nath06}).

The processes we have described here are clearly relevant for
determining the light element abundances. Much more work needs to  be
done before these processes can be combined with previous analyses to
put bounds on the model parameter space. Throughout this work we have
neglected the bound states of other light elements. The fraction of
bound states of D, T, \het,  \lisi and \lise are small owing to the
much smaller abundance of the respective light elements. However, they
could still have significant effects because of enhanced nuclear
cross-sections. The correct treatment of this problem including all
the cross-sections and bound states is beyond the scope of present
work.    

In summary, we have studied an aspect of the decay of charged
particles in the early universe. We looked at the
electromagnetic bound states of the charged particle with \hef nuclei
and considered the effects of the decay on the light element
abundances. We found that cosmologically relevant quantities of \lisi
could be produced as a result of these decays.

\section{Acknowledgements}
MK would like to acknowledge NSF grant PHY-0555689. AR would like
to acknowledge NSF grant PHY-0354993. 

\bibliography{cmb3}

\end{document}